\documentclass[reprint, amsmath,amssymb,aps,pra,superscriptaddress]{revtex4-1}
\usepackage{xspace,soul}
\usepackage[usenames,dvipsnames]{color}
\usepackage{amsmath} 
\usepackage{graphicx}
\usepackage{dcolumn}
\usepackage{bm}
\usepackage{physics}
\usepackage{mdframed}
\usepackage{qcircuit}
\usepackage{mathrsfs}
\usepackage{empheq}
\usepackage{psfrag,color}
\usepackage{subcaption}
\usepackage{todonotes}
\newcommand{\iso}[1]{\ensuremath{{^{#1}}}}
\newcommand{\tplus}{\ensuremath{^{3+}}\xspace}

\newcommand{\euform}{EuCl\ensuremath{_3\cdot}6H\ensuremath{_2}O\xspace}
\newcommand{\euformd}{EuCl\ensuremath{_3\cdot}6D\ensuremath{_2}O\xspace}

\newcommand{\isoeuform}{Eu$^{35}$Cl\ensuremath{_3\cdot}6H\ensuremath{_2}O\xspace}

\newcommand{\yso}{Y\ensuremath{_2}SiO\ensuremath{_5}\xspace}

\newcommand{\citeit}[1]{\cite{#1}}

\makeatletter
\def\Ddots{\mathinner{\mkern1mu\raise\p@
\vbox{\kern7\p@\hbox{.}}\mkern2mu
\raise4\p@\hbox{.}\mkern2mu\raise7\p@\hbox{.}\mkern1mu}}
\makeatother

\begin{document}
\title{Quantum processing with ensembles of rare earth ions in a stoichiometric crystal}

\date{\today}
\author{R. L. Ahlefeldt}
\email[Corresponding author:]{rose.ahlefeldt@anu.edu.au}
\affiliation{Centre for Quantum Computation and Communication Technology, Research School of Physics and Engineering, The Australian National University, Canberra 0200, Australia}
\author{M. J. Pearce}
\affiliation{Centre for Quantum Computation and Communication Technology, Research School of Physics and Engineering, The Australian National University, Canberra 0200, Australia}
\author{M. R. Hush}
\affiliation{School of Engineering and Information Technology,
University of New South Wales at the Australian Defence Force Academy, Canberra 2600, Australia}
\author{M. J. Sellars} 
\affiliation{Centre for Quantum Computation and Communication Technology, Research School of Physics and Engineering, The Australian National University, Canberra 0200, Australia}

\begin{abstract}
We describe a method for creating small quantum processors in a crystal stoichiometric in an optically active rare earth ion. The crystal is doped with another rare earth, creating an ensemble of identical clusters of surrounding ions, whose optical and hyperfine frequencies are uniquely determined by their spatial position in the cluster. Ensembles of ions in each unique position around the dopant serve as qubits, with strong local interactions between ions in different qubits. These ensemble qubits can each be used as a quantum memory for light, and we show how the interactions between qubits can be used to perform linear operations on the stored photonic state. We also describe how these ensemble qubits can be used to enact, and study, error correction.

\end{abstract}

\maketitle

\section{Introduction}\label{sec:introduction}
Small quantum processors can help make quantum networks practical and robust to errors. In a measurement-based quantum repeater, for example, a multi-qubit processor could purify entanglement \citeit{dur07, deutsch96, bennett96}, removing the errors caused by decoherence during photon transmission. Small processors could be used to generate the cluster states \citeit{raussendorf01} required for certain fault-tolerant communication schemes \citeit{barrett05}, or for blind quantum computation \citeit{fitzsimons17}. If sufficiently strong coupling can be generated between processors, scalable distributed quantum computing \citeit{nickerson13, jiang07a} will be possible. 

The physical systems suitable for making small quantum processors can look very different to those for full-size quantum computing, where scaling to large numbers of qubits is the primary concern. Small processors can prioritize high qubit interconnectivity and  strong qubit interactions. These properties suggest a system where the qubits are close together, such as spin clusters in solids. Strong optical coupling to these spin clusters is required, since most of the applications described above for small processors interface optically. Additionally, the operating wavelength and bandwidth should be matched to other network elements and the optical channel. One well-known example of such an optically addressable spin cluster system is nitrogen vacancy (NV) centers in diamond coupled to a random set of nearby \iso{13}C nuclear spins \citeit{childress06a,dutt07,neumann08}.   

In this paper we propose a new spin cluster system for generating small quantum processors: the rare-earth host ions surrounding a dopant in a rare-earth crystal (see Fig. \ref{fig:clusterconcept}). Tens of qubits could be resolved in such a system, and the short inter-ion distances mean strong interactions between qubits exist. The rare-earth ions have optically accessible hyperfine spin states, with long optical and spin coherence times at cryogenic temperatures. Rare-earth-based quantum repeater technology, such as quantum memories and quantum sources, is well developed, thus this system is automatically compatible with other key components of quantum networks.

This rare-earth system has three key advantages over other spin cluster systems, which we will briefly explain:  interconversion between optical and spin qubits, isolated two qubit gates, and a reproducible qubit layout. Considering the first advantage, the long optical coherence time means that an optical qubit can be created in addition to a spin qubit. Coherence on the optical qubit can be faithfully transferred to the spin qubit and back.  This means the system can be used to store and process photonic quantum information, required for most of the network applications of small quantum processors. This same property leads to the second advantage: the existence of both optical and spin qubits means two qubit gates on a pair of qubits can be isolated from other qubits in the system. The spin qubit is made the normal storage state for each bit of information, but optical interactions between neighboring ions are used to enact gates.  Since the spin qubits are unaffected by optical interactions, non-perturbing gates can be performed by transferring only the required spin qubits to the optical transition.

Finally, the qubit layout is reproducible because we use the set of host ions immediately surrounding a dopant, rather than the dopant ion itself. Thus, each cluster in the crystal is identical, and qubit-qubit distances are short. In schemes where dopants or impurities are used to form the cluster, such as NV centers with surrounding \iso{13}C impurities,  the cluster atoms are distributed randomly at low concentration, so each cluster is unique and qubit-qubit distances are large. 

A further consequence of the reproducible qubit layout is that, as well as using single clusters for processing, large ensembles of clusters can be used. Ensemble qubits can be easily read out optically, which is useful since rare-earth oscillator strengths are low and optical readout of single rare-earth ions is more challenging than for NV centers. 

We consider two applications of these small computing clusters. The first is quantum computing, computation directly on the spin qubits. In particular, these ensemble qubit systems are interesting for studying the effect of error correction protocols in the presence of real-world errors, which may be correlated between qubits. The second application is performing linear operations on photonic states. It has long been recognized that dilute rare earth crystals can perform the critical roles of a source and memory for non-classical photonic states in linear optics quantum computing (LOQC) applications \citeit{knill01a, obrien07}. In particular, the extremely long coherence times \citeit{rancic18,zhong15}, high memory efficiency \citeit{schraft16, hedges10}, ability to store multiple modes  \citeit{ferguson16,laplane16multiplexed} and to operate in the telecommunication band make rare earth crystals very appealing for implementing long range quantum repeater networks.  We will show that by moving from using dilute rare-earth crystals to an ensemble spin cluster system in a concentrated crystal, it is possible to implement LOQC with all the linear operations performed within the crystal itself. The approach avoids the inefficiency of repeatedly recalling and storing optical states to perform the linear operations. Further, because the output states remain in the memory, ready to be used as input to the next operation, it is possible to enact complex quantum circuits without the need for the complex spatial optical circuits used in more conventional approaches \citeit{obrien07}.

\section{The rare-earth spin cluster system}\label{sec:system}
\subsection{Overview}\label{sec:overview}

\begin{figure}[!ht]
\centering
\includegraphics[width=\linewidth]{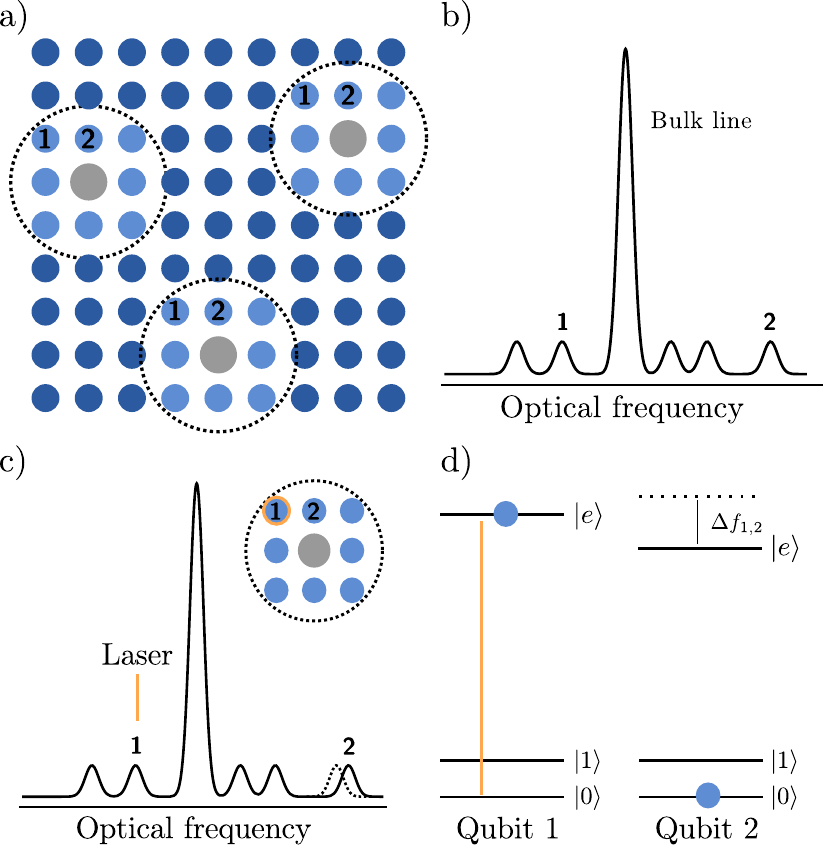}
\caption{\label{fig:clusterconcept} Concept for a rare-earth spin cluster quantum processor. a) Dopant ions (gray) in a rare-earth crystal cause localized strain, shifting the optical transition frequencies of surrounding host ions (blue). b) The resulting optical spectrum contains satellite lines, where each is caused by ions in a particular site relative to the dopant ion. Each optically resolvable satellite line is a potential frequency-addressable qubit. In the lowest symmetry (C$_1$) crystals, considered here, all surrounding sites are unique and lead to a unique satellite line. c), d) Two qubit gates are performed via the optical transition. Exciting one qubit (satellite line) shifts the optical frequency of nearby qubits.}
\end{figure}
The rare-earth spin cluster system we propose is shown in Fig. \ref{fig:clusterconcept}. We choose a host crystal stoichiometric in a rare-earth ion, like Eu$^{3+}$ or $^{167}$Er$^{3+}$,  that has at least three optically accessible hyperfine spin states. Spin clusters are created by lightly doping the crystal with a substitutional defect, such as another rare earth. The rare-earth ions surrounding this dopant serve as the spin cluster, each with a unique optical and hyperfine frequency allowing that ion site to be individually addressed. These frequency shifts arise because there is a size mismatch between the dopant and host ions, so the dopant distorts the crystal field at nearby lattice sites. Optical shifts of GHz to THz are common in rare-earth crystals \citeit{fricke79a,cone84,yamaguchi98}, observed as satellite lines in the optical transition spectrum of the crystal. The number of satellite lines depends, first, on symmetry of the crystal. Higher symmetry means less lines, so here we consider the ideal case of no symmetry (C$_1$). The number of these lines that are resolvable depends on the size of the optical shifts. Then, the number of resolved satellite lines determines the maximum number of qubits. Rare earth crystals often display 30 or more resolved lines (separated by more than the optical inhomogeneous linewidth)  \citeit{ahlefeldt13method, yamaguchi99}. More addressable satellite lines can be obtained by reducing the optical linewidth.

Single spin clusters or ensembles of spin clusters can both be used for computing. For ensemble qubits, we make use of the fact that each spin cluster in the crystal is identical: all ions in one position relative to the dopant have the same optical transition frequency (they contribute to the same satellite line), so these ions form one ensemble qubit. Crucially, each ion in one of these ensemble qubits also has the same separation to a partner ion in a second ensemble qubit, so interactions between all pairs of ions in two ensemble qubits are identical.

For either single instance or ensemble implementations, the unique optical frequency of each qubit in the spin cluster is normally used to address that qubit.  However, spin states are used for storage due to their long coherence times. Specifically, two hyperfine ground states are used as qubit levels $|0\rangle$ and $|1\rangle$, and a third ground state is used as a shelving level. This spin qubit can be readily transferred to an optical qubit for initialization, single and multi-qubit gate operations, and readout. 

For multi-qubit gates, we use the strong diagonal interaction on the optical transition, where the change in the wavefunction after exciting the first ion shifts the optical transition frequency of the second ion \citeit{ahlefeldt13precision}. This interaction is both strong and homogeneous: in \euform, we measured interactions between different pairs of satellite lines to be $\mathcal{O}$(10 MHz) with a homogeneity better than 1 kHz (the instrument limit)\citeit{ahlefeldt13precision}. Importantly, there is no significant interaction with hyperfine spin transitions, because the Stark shift of spin levels, $\mathcal{O}$(1 Hz.cm/V) \cite{macfarlane14} is much smaller than optical transitions $\mathcal{O}$(10-100 kHz.cm/V) \cite{macfarlane07}. Thus, isolated two qubit gates on pairs of qubits are possible:  optically exciting one qubit causes a unique shift in the optical transition frequency of surrounding qubits, but does not affect the information stored in spin qubits. It is only when a second qubit is transferred to the optical transition that a  two-qubit gate can be enacted.

Because all ions in a spin cluster are close together, the qubits are strongly interconnected by the frequency shift interaction. As we discuss in Section \ref{sec:materials}, $\mathcal{O}$(10) qubits are strongly interconnected in the Eu\tplus example system we consider later. Stronger interactions could be expected for Kramers ions like Er\tplus. Thus, gates can be enacted directly between any pair of qubits unless the computing system is large. Since the interaction strength between any pair of qubits is typically unique and qubit interactions sum linearly, the strong interconnection means that gates can also be enacted between three or more qubits in this system.

Rare earth ensembles have previously been proposed for quantum computing \citeit{ohlsson02,longdell04a,yan13}. All those proposals used crystals where the optically active ion was a randomly distributed dopant, in contrast to the stoichiometric crystals considered here. This choice leads to differences in how the computer operates and scales, which are discussed in Section \ref{sec:comparison}. However, the techniques for qubit initialization, gate operations, and readout are very similar, and so we will only briefly explain these here.

\subsection{Initializing qubits}\label{sec:initialise}
The hyperfine ground states of rare-earth ions are split by MHz to GHz, which means that the two computing states $\ket{0}$ and $\ket{1}$ are nearly equally populated even at cryogenic temperatures. Therefore, initializing a qubit requires transferring the qubit to a well defined state, e.g. $|0\rangle$. For a single-instance spin cluster, this is simple to achieve using standard optical pumping techniques. For instance, for a system with three ground states ($|0\rangle$, $|1\rangle$ and $|aux\rangle$) and one shared optical state $|e\rangle$ it is sufficient to use two optical fields tuned to the transitions $|1\rangle-|e\rangle$ and $|aux\rangle-|e\rangle$ to drive the atom into $|0\rangle$. 

For ensemble qubits, the initialization process depends on the optical inhomogeneous width of the satellite line used as a qubit. There is a maximum permissible inhomogeneous width for the prepared qubit, because the optical pulses used for gates must be able to drive all ions in the qubit equally. As discussed in more detail in Sec. \ref{sec:materials}, the maximum permissible width is dependent on the hyperfine splitting and the available optical Rabi frequency. If the ensemble inhomogeneous width is smaller than this maximum width, the optical pumping process described above for single instance qubits can be used.

If the optical inhomogeneous linewidth is larger than this maximum permissible width, extra spectral holeburning steps are required to initialize the qubits. The technique is similar to that demonstrated for ensemble qubits in rare earth doped systems \citeit{nilsson04a,sellars03}.  A spectral trench is burnt in the line, transferring the population to the shelving state $\ket{aux}$, and then holeburning at other frequencies creates a narrow feature -- the qubit feature -- in the middle of this trench. The frequencies are chosen so that the ions making up the qubit feature are also initialized into a known state, initializing the qubit. When using this type of qubit feature, the residual population in the trench must be low to avoid adding noise. This is typically straightforward in rare earth crystals, because the hyperfine lifetime is long and the system can be chosen such that oscillator strength on the holeburning transition is non-zero. 

\begin{figure}[!ht]
\includegraphics[width=\linewidth]{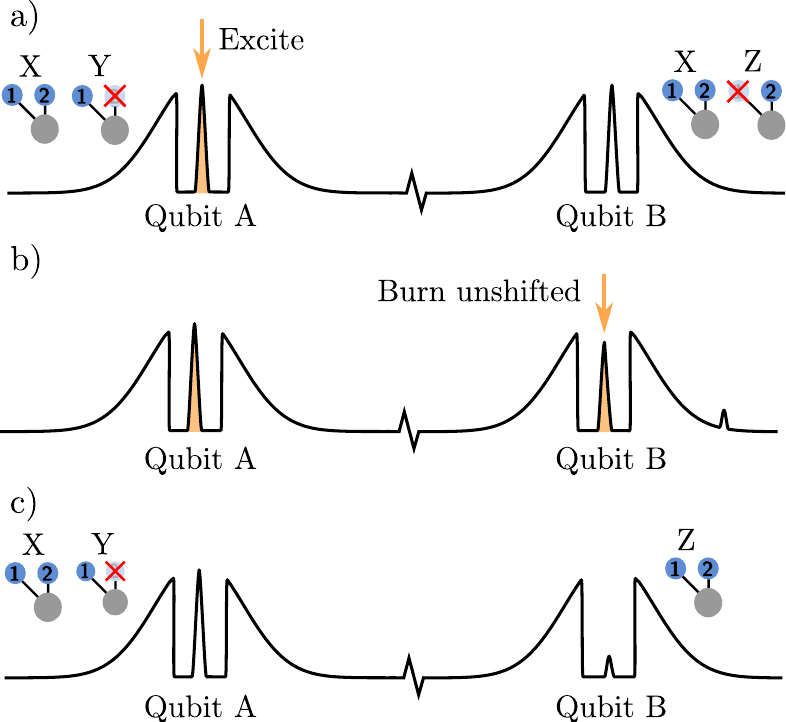}
\caption{\label{fig:distill} Distillation process for ensemble qubits produced by spectral holeburning \citeit{longdell04a}. (a) Holeburning in two satellite lines produces two qubit features. Three types of spin clusters can contribute to the qubit features: for qubit A, type X (both ions in the qubit feature) and Y (only ion 1 in the qubit feature), for qubit B, type X and Z (only ion 2 in the qubit feature). Types Y and Z have to be discarded since no interaction between qubits is possible. This is achieved by exciting qubit A. The optical interaction between qubits shifts all type X clusters, (b). By repeatedly exciting qubit A and then the unshifted portion of B (type Z clusters), the unshifted portion is holeburnt to a shelving state.  This leaves only type X clusters in qubit B, (c). Once qubit B has been distilled in this way, the process must then be repeated exciting qubit B to discard type Y clusters and distil qubit A.}
\end{figure}
This spectral holeburning process can be repeated on other satellite lines to create additional ensemble qubit features. However, this process requires discarding a large proportion of ions in each satellite line. Thus, only a small proportion of the corresponding spin clusters have an ion in every qubit feature, and a distillation process is required to ensure that ions in one qubit feature can interact with ions in another qubit feature \citeit{longdell04a}.

Qubits are distilled using the strong optical frequency shift interaction that will be used for two-qubit gates. The process is shown in Fig. \ref{fig:distill}. For each pair of qubit features, one qubit feature is completely excited, which will shift the frequency of those ions in second qubit feature that neighbor an excited ion. In the materials we will consider, the shift is much larger than the qubit feature width, so any unshifted ions can simply be holeburnt to the shelving state, discarding those spin clusters. The process is then repeated in reverse, exciting the second qubit feature, and then on every pair of qubit features in the system. This type of process was first described and demonstrated for ensemble qubits in rare earth doped crystals, although with the complication that the interactions were smaller than the qubit feature width \citeit{longdell04a}. 

This type of distillation process has an additional use. Here we consider using ideal low symmetry crystals as hosts,  but crystals with symmetry above C$_1$ can also be used. However, each satellite line will  consist of multiple sites with different interaction strengths. All but one of these equivalent sites in each satellite line will need to be discarded so that each pair of qubits has a single interaction strength, which can be achieved during the distillation process by burning away ions with undesired shifts, along with the unshifted ions.

\subsection{Gates and readout}\label{sec:gates}
Once qubits are prepared and initialized by one of the above techniques, single qubit gates are enacted using a series of optical pulses \citeit{rippe05}. Arbitrary rotations are possible by modifying the pulse length and phase. For example, a NOT gate can be enacted by three $\pi$ pulses at different frequencies: first on $|0\rangle-|e\rangle$ to transfer the spin qubit to an optical qubit, then on $|1\rangle-|e\rangle$ to enact the gate, and finally on $|0\rangle-|e\rangle$ to transfer the optical qubit back to a spin qubit. The quality of single qubit gates depends, principally, on the time the gate takes relative to the optical coherence time, and whether the pulses off-resonantly drive undesired transitions. 

\begin{figure}
\centering
\begin{subfigure}{0.48\linewidth}
\includegraphics[width=0.9\textwidth]{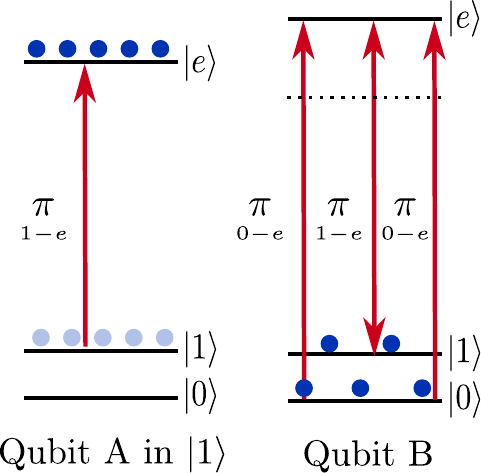}
\caption{\label{fig:cnotseq}}
\end{subfigure}
\begin{subfigure}{0.48\linewidth}
\includegraphics[width=0.9\textwidth]{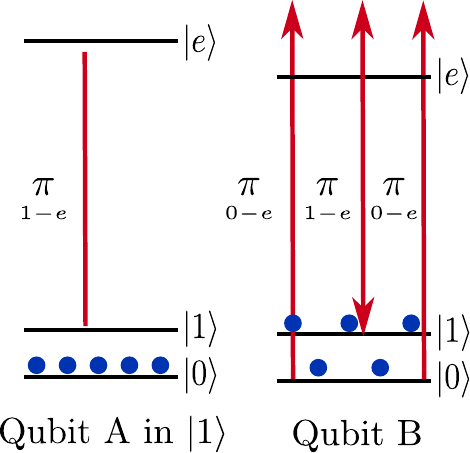}
\caption{\label{fig:cnotseq}}
\end{subfigure}
\begin{subfigure}{0.9\linewidth}
\includegraphics[width=\textwidth]{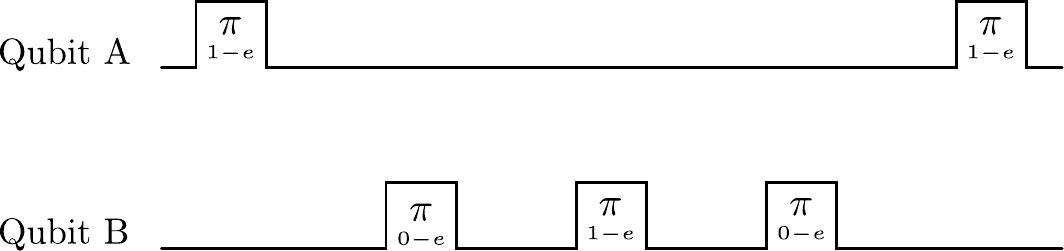}
\caption{\label{fig:cnotseq}}
\end{subfigure}
\caption{\label{fig:cnot} Illustration of a CNOT gate in a rare earth ensemble system. The target is qubit B, the control is qubit A. Two cases are displayed. In (a), qubit A is in $\ket{1}$, is excited, and the NOT gate on B flips the state. In (b), qubit A is in $\ket{0}$, is not excited, and the NOT gate is non-resonant with qubit B so it is left unchanged. (c) corresponding pulse sequence.}
\end{figure}
The same optical pulses are used for two qubit gates. For example, as illustrated in Fig. \ref{fig:cnot} a CNOT gate can be enacted by exciting the control qubit with a $\pi$ pulse on $|1\rangle-|e\rangle$ to transfer it to an optical qubit, and then applying a single qubit NOT gate on the target qubit at the shifted frequency. If the control qubit was in  $|1\rangle$, it is excited and the target qubit shifts into resonance with the pulses of the NOT gate due to the interaction between the qubits, so it is flipped. If the control qubit started in $|0\rangle$, it is not excited and the NOT gate does not proceed, since it is not resonant with the target qubit. A three qubit controlled-CNOT (CCNOT) gate, where the target qubit is flipped only two control qubits are in $\ket{1}$, can be enacted with a similar method. Both control qubits are driven by $\pi$ pulses on $|1\rangle-|e\rangle$, and then a shifted single qubit NOT gate is applied to the target.  This simple implementation for two and three qubit gates requires that the target qubit is completely non-resonant with the NOT gate unless the control qubit is excited, but equivalent gates exist if this is not the case \citeit{longdell04a}.

Readout of each qubit is again performed via the optical transition. Readout of ensembles of rare earth ions is straightforward using standard photon echo methods. For single-instance spin clusters, readout is the most challenging part of the computation as rare-earth oscillator strengths are low, but methods are being developed to read out single rare-earth ions using optical up-conversion \citeit{kolesov12,siyushev14}, and strongly coupled optical cavities \citeit{zhong18}. Readout of the quantum state of a single Pr$^{3+}$ spin has recently been achieved \citeit{xia17}, showing that addressing spin states of single rare earth clusters is possible. We also have the flexibility to choose the dopant ion, which generates the spin cluster, to have a high oscillator strength. This dopant ions can then be used as a readout ion, similar to what has been suggested for doped rare-earth crystals \citeit{yan13}. 

\subsection{Comparison with other frequency addressed computing systems}\label{sec:comparison}
This stoichiometric rare-earth system resembles other frequency-addressed qubit systems, but has some unique features. Here, we explain these in more detail.

The most well-known frequency addressed quantum computing system is liquid state NMR \citeit{cory97,gershenfeld97, yamaguchi99a}. This approach uses ensembles of identical molecules for computing, where the different nuclear spins on each molecule form the frequency addressed qubits.  Liquid state NMR was initially heavily studied, but ultimately this approach was  limited by two key drawbacks. The ensemble qubits cannot be initialized into a pure state, and isolated two-qubit gates are not possible, i.e. a two qubit gate applied between two qubits will affect the state of every other qubit in the system. These problems arise because NMR qubits are two-level systems near room temperature, split only by a small Zeeman effect. Initialization is not possible because the levels are nearly equally populated and it is not possible to pump the qubits into one state. Isolated gates between pairs of qubits are prevented because the system has only two levels, so driving one qubit must necessarily affect all surrounding qubits, not just the one other qubit participating in the gate. Because of this, NMR uses complex pulse sequences to apply the desired two qubit gate and to 'reverse' the effect of the gate on other qubits.

We stress that these problems do not apply to quantum computing systems based on rare earth ions in solids. The reason is that we use cryogenic temperatures and an optical transition coupled to the spin transition for initialization and gates. The spin levels that form the computing states are typically equally populated even at low temperatures. However, the entire ensemble can be optically pumped into a single, pure state with lifetimes as long as weeks  at cryogenic temperatures\citeit{konz03}. As explained in Section \ref{sec:overview}, the optical transition also allows isolated gates. While rare earth qubit systems do not share the main disadvantages of NMR, they do share the main advantages. Compared to single instance approaches to quantum computing, it is easier to create small qubit systems since many frequency resolved satellite lines exist. Sophisticated pulse sequences for manipulating the quantum state of nuclear spins have existed for many years, and were further developed by the NMR quantum computing community. 

More recently, quantum computing has been proposed using single NV centers coupled to a random distribution of nearby \iso{13}C \citeit{childress06a,dutt07,neumann08}. In this system, single qubit gates can performed directly on the \iso{13}C nuclear spins, but two qubit gates must be enacted via the NV electron spin, different to rare earth systems where qubits are directly connected. Similar to rare earth systems, NV systems are initialized and read out via the optical transition, but  the short coherence time means that you cannot exchange quantum information from the spin qubit to an optical qubit, which is possible in the rare earths.

The final frequency addressed system we will consider is single-instances and ensembles of rare-earth dopants in crystals \citeit{ohlsson02,longdell04a,yan13}. These approaches are very similar to what we propose here.  Single qubit gates have been demonstrated in rare earth doped crystals with gate fidelities of up to 96\% observed \citeit{longdell04, roos04, rippe08}.  The main difference been those doped systems and our stoichiometric approach is that, like NV centers, doped crystals use a spin cluster made up of atoms randomly distributed in the crystal. This prevents the use of large ensembles of spin-clusters: since the positions of spins is different in each cluster, the qubit-qubit interaction strengths, and the qubit frequencies themselves, are highly inhomogeneous. It is possible to select out a sub-ensemble that is more homogeneous in frequency and interaction strength, but the ensemble size is typically small, and becomes exponentially smaller as more qubits are added to the system. 

Only a low-fidelity two qubit gate has been experimentally demonstrated in a doped rare-earth ensemble due to the inhomogeneity in the ensemble \citeit{longdell04a}. To avoid this problem, a single instance approach has been proposed where the rare earth ion is coupled to a bus qubit for readout \citeit{yan13}, but gates have yet to be demonstrated. In the scheme we propose here, ensembles can be used because the inhomogeneity in the separation of the spin clusters is removed by using host rather than dopant ions. This greatly improves the exponential scaling of the ensemble size with qubit number, and in the limit of very narrow optical inhomogeneous linewidths, the limit is removed altogether.

\section{Applications of rare earth ensemble qubits}
\subsection{Quantum computing and error correction with spin qubits}
A key advantage of the quantum computing scheme that we proposed above is that it is simple to create small qubit systems. The number of qubits that can be used at once depends on the number of ions in each ensemble qubit after the set of qubits has been distilled and initialized, compared to the minimum number of ions the readout method can detect. As shown in Section \ref{sec:materials}, five qubit systems are possible in current materials, and the path to scaling to larger numbers of qubits is a matter of materials optimization. Here, we consider what could be done with systems of three to ten qubits, likely achievable with fairly modest improvements in the materials used. More qubits will greatly enhance the usefulness of this system: since qubits in this system are highly connected (Sec. \ref{sec:materials}) and arbitrary two and three qubit gates are possible (Sec. \ref{sec:gates}), this system will be suitable for a wide variety of fault-tolerant codes once the qubit number is increased.

An immediate application is the study of errors and error correction. The ability to correct errors during a computation is a fundamental capability required of any quantum computing system. This spin cluster system could furnish exactly the required small qubit system to test error correction protocols: not only are systems of five qubits possible now, the ensemble approach means that the experimental implementation is simple. The sample need only be maintained at liquid helium temperature and moderate magnetic field, and high fidelity gates and readout can be achieved with standard optical pulses applied to the entire sample. 

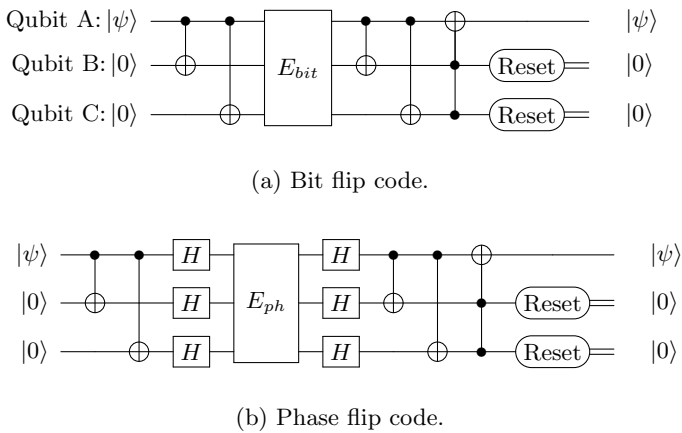
\begin{figure} 
\begin{subfigure}{\linewidth}
\centering
\begin{equation*}
\Qcircuit @C=1em @R=.7em {
&& & \lstick{\textrm{Qubit A:}  \ket{\psi}} &\ctrl{1}  &\ctrl{2} &\multigate{2}{E_{bit}} &\ctrl{1} &\ctrl{2} &\targ     &\qw	&\qw	 & \rstick{\ket{\psi}}\\ 
&& &\lstick{\textrm{Qubit B:} \ket{0}}    &\targ     &\qw      &\ghost{E_{bit}}           &\targ    &\qw       &\ctrl{-1} & 	\measure{\mbox{Reset}} & \cw & \rstick{\ket{0}}\\ 
&& &\lstick{\textrm{Qubit C:}  \ket{0}}    &\qw       &\targ    &\ghost{E_{bit}}		     &\qw	   &\targ     &\ctrl{-2} & \measure{\mbox{Reset}} & \cw & \rstick{\ket{0}}
}
\end{equation*}
\caption{Bit flip code. \label{fig:bit}}
\end{subfigure}
\begin{subfigure}{\linewidth}
\centering
\begin{equation*}
\resizebox{.9\linewidth}{!} 
{
\Qcircuit @C=1em @R=.7em{
\lstick{\ket{\psi}}  &\ctrl{1} &\ctrl{2} &\gate{H} &\multigate{2}{E_{ph}} &\gate{H} &\ctrl{1} &\ctrl{2} &\targ & \qw & \qw & \rstick{\ket{\psi}}\\
\lstick{\ket{0}}  &\targ &\qw &\gate{H} &\ghost{E_{ph}} &\gate{H} &\targ & \qw & \ctrl{-1} & \measure{\mbox{Reset}} & \cw & \rstick{\ket{0}}\\
\lstick{\ket{0}} &\qw &\targ &\gate{H} &\ghost{E_{ph}} &\gate{H} &\qw &\targ & \ctrl{-2} & \measure{\mbox{Reset}} & \cw & \rstick{\ket{0}}\\
}
}
\end{equation*}
\caption{Phase flip code. \label{fig:phase}}
\end{subfigure}
\caption{\label{fig:protocol} 3-qubit quantum error correction protocols for ensemble qubits, which preserve a data qubit A. \subref{fig:bit}  In the bit flip code, two CNOT gates are used to prepare an entangled state. After allowing time for an error to occur, the CNOT gates are reversed,  any error on qubit A is corrected (CCNOT), and qubits B and C are reset to $\ket{0}$. \subref{fig:phase} In the phase flip code, Hadamard gates $H$  convert any phase flip error into a bit flip error that is then corrected.}
\end{figure}
Ensemble error correction protocols have previously been applied in liquid state NMR \citeit{cory98, knill01}. These protocols work differently to those for single instance computing approaches. Both involve using entangling gates to encode the original qubit state onto multiple qubits. In single instance the next step is to perform projective measurements to identify the error, and apply a gate to correct it, restoring the encoded logical qubit state. In ensembles each instance can have different errors and so projective measurements are not useful. Instead the additional qubits are treated as ancillaries: a series of gates disentangles the qubits and transfers any error on the original qubit onto these ancillary qubits, which are then reset. Additional error correction steps involve reapplying the entire error correction sequence, including the encoding step, in contrast to the single instance approach. In Fig. \ref{fig:protocol}, we give examples of 3-qubit codes for correcting single bit flip or phase flip errors. Larger qubit codes, such as the seven-qubit Steane code or the nine-qubit Shor code can be implemented with the same modifications: ancilla qubits are reset, not measured, and the qubit encoding step is repeated each time the error correction protocol is run.

This rare earth spin cluster can be engineered to investigate different error environments using these error correction protocols. In particular, understanding the correlation of errors is important because error correction protocols assume uncorrelated errors \citeit{preskill13}, but in the real systems proposed so far for quantum computing errors are often highly correlated \citeit{edmunds17}.

A rare earth spin cluster can be manipulated to have different correlations. This is possible because the spin cluster has many potential qubits, any optically resolvable satellite line, but we can only distill, and thus use for quantum computing, a smaller number of these at any one time. Thus, for the computing cluster we are free to choose a set of these that has the characteristics we want. Not only this, but by sequentially defining different computing clusters, we can characterize the errors, interactions, and correlations that would exist in a much larger qubit system. For instance, if we can distill five qubit systems, we could characterize all pair-wise correlations and interactions in a ten qubit system by characterizing 21 different five-qubit systems. Then, we could accurately predict the performance of a ten qubit system once the inhomogeneous linewidth of the material is lowered sufficiently to allow such a system to be distilled.

Two main types of natural errors exist in rare earth crystals: spin-lattice interactions with phonons, and  dephasing due to the fluctuating magnetic field created by the bath of nuclear spins. Spin-lattice relaxation causes an arbitrary error which is completely uncorrelated, while spin-bath interactions are a pure phase error, and partially correlated between two qubits depending on their separation from each other, and on the magnetic field. Spin-lattice interactions are largely only temperature dependent, but spin-bath interaction can be manipulated to study different error regimes. For instance, in most magnetic fields, spin-bath interactions are the major error source, but in most rare earth materials, there exist fields where a rare earth spin transition goes through a turning point in frequency \cite{fraval04}, and the coupling to the nuclear spin bath is greatly reduced. Likewise, the spatial dependence of correlations can be probed by choosing the separation between qubits. 

It is also possible to artificially create errors of a particular probability and type, including those that do not naturally occur in rare earth crystals, such as bit-flip errors. To do this, we would utilize one of the extra satellite lines that has not been distilled into a qubit. Exciting some proportion $C$ of ions in this line would shift the optical transition frequencies of the same proportion of the computing spin clusters, randomly. An optical bit-flip gate then applied to the shifted frequency of one of the qubits would add a random bit flip error to the qubit with a probability $C$. Any optical gate or sequence of gates could be applied instead of a bit flip gate, and multiple satellite lines could be excited. This approach would provide great flexibility in generating different errors to test error correction protocols.

\subsection{Linear processing of photonic states}
So far, we have considered using this ensemble spin system for quantum computing with the spin qubits themselves. However, the system also furnishes an optical qubit coupled to the spin qubit. This has two implications: first, each ensemble qubit can be used as a quantum memory for a photonic state, and second, the existence of gates between ensemble qubits means that operations can be performed on the photonic qubit.  We will show that the various operations used in linear optics quantum computing can be performed using these spin cluster systems, which means these ensemble spin cluster systems can perform tasks in a quantum network currently proposed for linear optics, such as entanglement swapping and distillation. There is already work on doing these linear operations in Raman-style quantum memories in atomic gases \cite{campbell14, lee14, mazelanik19}. At the end of this section, we will discuss how that approach differs from the one we describe here.

In the spin-cluster system, the first step in any manipulation of a photonic qubit is to transfer it to a spin qubit on one of the satellite lines of the crystal, which is achieved via a quantum memory protocol. Rare earths have been heavily studied for quantum memories, with nearly all that work on low concentration dopant ions in an inactive host, functionally equivalent to a single satellite line here. In a quantum memory, a photonic state is absorbed by a spectrally and often spatially tailored ensemble of rare earth ions, in such a way that the state can be recalled later on demand.  While the photonic state is in the memory, it is stored as an entangled state across the ensemble. If the photonic state is a single photon, the stored state is a superposition of each ion having absorbed the photon:
\begin{equation}
\ket{\psi} = \frac{1}{\sqrt{N}}\sum_{n=1}^N e^{i\bm{k}\cdot\bm{x_n}}\ket{g_1g_2...e_n...g_{N-1}g_{N}}
\label{eq:memorystate}
\end{equation}
where $e^{i\bm{k}\cdot\bm{x_n}}$ is a spatial phase factor given by the ion's position in the ensemble and wavevector of the light $\bm{k}$, which determines a phase matching condition for recall of the light. This state is quite different from the state used for quantum computing, where each spin cluster has its own local copy of the state to be computed, and all computations are performed within the cluster, but done in parallel across the ensemble. 

This leads to differences in the operations that can be performed on photonic qubits compared to spin qubits. Since the photonic state is effectively spread across the very large ensemble, the interaction between two states stored on two different ensemble qubits is weak, although the interactions between the individual ions in each spin cluster are strong.  This means that deterministic two-qubit gates on stored photonic states are not possible. To make this point completely clear, imagine 1 million ions contribute to each ensemble qubit. If each ensemble qubit absorbs a single photon, the probability that it was the two ions in \textit{one} spin cluster that absorbed the two photons is negligible, 10$^{-12}$. But a deterministic two qubit gate on the photonic state requires changing the state of one photonic qubit \textit{conditional} on the state of a second photonic qubit, so it has the same, negligible, chance of success. Only for that negligible probability state does the first stored qubit ``see'' the state of the second stored qubit.

  However, it is possible to use the spin-qubit gates to perform \textit{linear} operations on photonic states (the operation on one state is independent of the other state). These operations  include single-qubit phase gates, a SWAP gate, which switches the modes of two photonic qubits, and a beamsplitter operation that mixes the modes. To illustrate these operations, consider we have two photonic states at frequencies A and B: $c_1\ket{0}+c_2\ket{1}$ and $d_1\ket{0}+d_2\ket{1}$, where $\ket{0}$ is the vacuum state. $\ket{1}$ could be any small-photon-number Fock state or a weak coherent state (we require only that the photon number is much smaller than the size of the ensemble qubit $N$), but for simplicity we assume it is a single-photon state. The two photonic states are stored on ensemble ionic qubits A and B  using a quantum memory technique. Ignoring spatial phase, the stored quantum state on the two ensemble qubits is:
\begin{equation}
\ket{\psi}=\frac{1}{N}\sum_{j}^N\sum_{k}^N \left(c_1\ket{\mathbf{0}}_A+c_2\ket{\mathbf{1}_j}_A\right)\left(d_1\ket{\mathbf{0}}_B+d_2\ket{\mathbf{1}_k}_B\right)
\label{eqn:state}
\end{equation}
where we have defined $\ket{\mathbf{0}} = \ket{0_10_2...0_{N-1}0_N}$ and $\ket{\mathbf{1}_j} = \ket{0_10_2...1_j...0_{N-1}0_N}$, that is, a single excitation on the $j$-th ion in the ensemble qubit. We will illustrate the effect of a SWAP gate on this state. A well-known way to achieve this gate is with three consecutive CNOT gates on alternating qubits, each enacted by the pulse sequence shown in Fig. \ref{fig:cnot}. The first CNOT gate is applied on qubit B, controlled on A, after which the system state is:
\begin{align}
\ket{\psi'}=\frac{1}{N}&\sum_{j}^N\sum_{k\neq j }^N c_1\ket{\mathbf{0}}_A\left(d_1\ket{\mathbf{0}}_B+d_2\ket{\mathbf{1}_k}_B\right)\nonumber\\ &+c_2\ket{\mathbf{1}_j}_A\left(d_1\ket{\mathbf{1}_j}_B+d_2\ket{\mathbf{1}_k}_B\right)
\end{align}
 We note that this state does not correspond to a CNOT gate on the photonic state -- to do that, we would require that the gate turns states of the form $\ket{1}_A\ket{1}_B$ into states of the form $\ket{1}_A\ket{0}_B$. The only way this transformation could be achieved would be if the A and B ions in a single cluster absorbed the two photons, but  since the ensemble is large the chance of this is negligible. It is for this same reason that we ignored the $k=j$ term. 
 
The second CNOT gate is applied on qubit A, giving the state:
\begin{align}
\ket{\psi''}=\frac{1}{N}&\sum_{j}^N\sum_{k\neq j }^N c_1d_1\ket{\mathbf{0}}_A\ket{\mathbf{0}}_B+c_1d_2\ket{\mathbf{1}_k}_A\ket{\mathbf{1}_k}_B\nonumber\\
&+c_2d_1\ket{\mathbf{0}}_A\ket{\mathbf{1}_j}_B+c_2d_2\ket{\mathbf{1}_k}_A\ket{\mathbf{1}_j\mathbf{1}_k}_B
\end{align}
where $\ket{\mathbf{1}_j\mathbf{1}_k} = \ket{0_1...1_j...1_k...0_N}$ indicates both the $j$-th and $k$-th ions in the ensemble qubit are excited. The final CNOT on qubit B gives:
\begin{equation}
\ket{\psi}=\frac{1}{N}\sum_{j}^N\sum_{k}^N \left(d_1\ket{\mathbf{0}}_A+d_2\ket{\mathbf{1}_j}_A\right)\left(c_1\ket{\mathbf{0}}_B+c_2\ket{\mathbf{1}_k}_B\right)
\label{eqn:finalstate}
\end{equation}
and comparison with Equation \eqref{eqn:state} shows that we have performed a SWAP operation, not only on the ionic ensemble qubits but also on the stored photonic states. 

Other, more efficient, implementations of the SWAP gate are also possible in this system. For instance, we can perform a SWAP with the following seven-pulse sequence (the CNOT method requires 15 pulses): 
\begin{equation}
\textrm{SWAP} = \pi_{1-e}^A\,\pi_{1-e}^B\,\pi_{0-e_{\Delta}}^{A}\,\pi_{0-e_{\Delta}}^B\,\pi_{0-e_{\Delta}}^A\,\pi_{1-e}^A\,\pi_{1-e}^B
\end{equation}
where superscripts are the qubit driven, subscripts the transition driven, and $\Delta$ indicates that the pulse is controlled: the transition is driven at the frequency shifted by the optical interaction with qubit B. This sequence does lead to an error when both A and B ions in a cluster absorbed the photonic state, but as explained above that possibility makes a negligible contribution to the ensemble state for a large ensemble. The sequence can easily be generalized to an arbitrary beamsplitter gate by replacing the central $\pi_{0-e_{\Delta}}^B$ pulse with a different rotation. For example, a $\frac{\pi}{2}$ pulse gives the equivalent of a 50/50 beamsplitter on the photonic states.

Now, we also need to be able to recall these stored states. Equation \eqref{eq:memorystate} showed that the initial storage encodes spatial phase on the ensemble , $e^{i\bm{k}\cdot\bm{x_n}}$, and any subsequent control pulses at the same frequency will have the same phase variation. Quantum memory schemes are designed to meet phase matching conditions so that this phase variation is removed in the recall process. Here, though, we have swapped photonic states initially stored at two different frequencies, so they will have slightly different spatial phase.  As long as the variation is small, the states can still be recalled efficiently. This then means we require that the wavelength of the difference frequency between the two ensemble qubits is small compared to the crystal length. For ensemble qubits separated in optical frequency by 10 GHz, for example, phase errors would become important once the crystal size was $>1$~cm. Quantum memory experiments typically use crystals smaller than 1~cm, so this effect can be ignored. Larger crystals can also be used if smaller separations between the qubits are chosen. 

The above example has shown that we can use this system to perform SWAP operations on stored photonic states, one of the key operations for linear optics quantum computing. We also need to perform single qubit rotations, and these are not possible using the single-rail encoding of the photonic state (in photon number) given above. We need a different encoding scheme. Time-bin encoding is straightforward in these systems, but suffers the same problem that it is not clear how to perform single qubit gates on time-bin qubits \citeit{kok07}. Instead, we need a dual rail encoding \citeit{kok07,obrien07}. Commonly, states are encoded in orthogonal polarization or spatial modes, but neither is well suited to our ensemble spin-cluster system. Spatial encoding is not possible because we rely on atomic-scale interactions to perform gates, so all qubits must exist in the crystal in the same spatial mode. Polarization encoding is possible, but would require a material with polarized transitions and a very specific crystal symmetry, with two sets of sites with orthogonal polarizations. This strongly limits the number of usable materials.

An alternate dual-rail encoding that would work in our ensemble spin system is frequency-encoding: qubits are encoded across two frequency modes, where each mode is resonant with one satellite line. The qubit states are $\ket{0}_q = \ket{1}_A\ket{0}_B$ and $\ket{1}_q = \ket{0}_A\ket{1}_B$, and can be stored in the memory in the same way as the single-rail photonic state in Equation \eqref{eqn:state}. Then, the SWAP gate in Equations \eqref{eqn:state}--\eqref{eqn:finalstate} corresponds to a single-photonic-qubit NOT gate, and other single qubit rotations are achieved with similar operations. To apply a SWAP gate to two photonic qubits, $\ket{\psi}_{q1}$ stored on ensemble qubits A and B, and $\ket{\psi}_{q2}$ stored on ensemble qubits C and D, it is sufficient to apply SWAP gates sequentially, first on A and C, and then on B and D.

Implementing linear optics quantum computing in a quantum memory without introducing significant error in the stored state requires careful choice of the pulse sequences used for any of the gates described above. Storing a photonic state with high efficiency requires a high optical depth on the ensemble qubit. But if the optical pulses required to enact gates on the stored qubit are applied on high optical depth transitions, they will cause two problems. First, the pulses will be distorted as they travel through the crystal, causing errors. Second, the pulses will cause substantial optical inversion of the atoms, which leads to amplified spontaneous emission noise, degrading the stored quantum state. Thus, all optical pulses should be applied on low optical depth transitions, that is transitions that are either detuned or nearly unpopulated. This is straightforward for weak stored photonic states: the excitation in the ensemble is low, and after the state is transferred to the computing levels, the population is mostly in $\ket{0}$. When processing a dual-rail photonic state stored across two ensemble spin qubits, there is never any need to transfer the state into $\ket{1}$. Thus, the $\ket{1} -\ket{e}$ transition always has low optical depth, and low optical depth can be obtained on the $\ket{0} -\ket{e}$ transition by choosing pulses controlled on $\ket{1}-\ket{e}$ transitions. The implementation of CNOT gates described in Section \ref{sec:gates} meets these requirements, and in fact, any of the gates described above can be designed to meet these requirements.

In summary, both quantum memories and linear operations are needed for quantum networks, and this system enables an implementation of linear optics quantum computing built into a quantum memory. Linear operations have previously been demonstrated in Raman-type atomic ensemble quantum memories, (e.g. \cite{campbell14, lee14, mazelanik19}), and we will briefly compare that approach to the one here. In those memories, linear operations like beamsplitter gates on the stored spin states are achieved by either off-resonantly driving multiple internal states of the atomic ensemble to interfere spin waves stored on different frequency modes \cite{campbell14, lee14}, or far-off resonantly driving one internal state with spatially and temporally shaped fields to interfere spin waves stored on different temporal-spatial modes via an AC Stark shift \cite{mazelanik19}. In those memories, then, the linear operations are performed on optical states that are \emph{propagating} through the memory. Thus, for example, in \cite{campbell14}, operations can be performed only during the read-in or read-out stages of the memory. In the approach we describe here, linear operations are performed by resonantly transferring the spin qubit  to a \emph{stationary} optical qubit and using optical interactions on this non-propagating state. Thus, many operations can be executed sequentially without ever recalling the memory, which should allow complex circuits to be executed while maintaining a high storage efficiency. Another useful feature is that operations on two photonic qubits can be performed without perturbing other qubits in the system, unlike in Raman-style schemes where all off-resonant drive fields necessarily perturb all qubits in the system.  The ability to swap qubits between frequency modes in this system also provides a means of wavelength division multiplexing.

\section{Material considerations} \label{sec:materials}
The previous sections described two uses of rare earth ensemble spin clusters: using the spin qubits for quantum computing, particularly error correcting, and performing linear processing of photonic states. This section discusses how ensemble qubits would be implemented in real materials, and how the material properties affect the number of qubits, and the quality of the computing operations.  For processing photonic states, we will assume a gradient echo memory (GEM) protocol is used to transfer the state onto the atomic ensemble, since this requires no extra preparation of the ensemble qubits, but other protocols can be used with suitable preparation.

As we outlined in Section \ref{sec:system}, the basic system we need to create a suitable multi-qubit ensemble cluster is a crystal stoichiometric in one rare earth ion and doped with a second to produce resolved satellite lines. The ions in these satellite lines must have an optical frequency-shift interaction between them but no other large interactions (such as energy transfer). 

One of the most important considerations is the optical inhomogeneous linewidth. The prepared qubit needs a spectral width smaller than both the hyperfine splittings and the optical Rabi frequency to minimize gate errors caused by off-resonant driving of the qubit or off-resonant excitation of other levels. The ideal material has a natural linewidth in this limit. Given a hyperfine splitting of 10-100 MHz, the maximum allowable linewidth would be $\mathcal{O}$(1 MHz).

In a broader material, an artificially narrow qubit can be prepared by holeburning as described in Sec. \ref{sec:initialise}. However, this then requires a distillation process to define an ensemble of spin clusters where all the qubits interact. Distillation has the disadvantage that it decreases the ensemble size exponentially with the number of qubits $n$ as 
$\left(\frac{\Gamma_q}{\Gamma_{inh}}\right)^{n-1}$, where $\Gamma_q$ is the width of the prepared qubit, and $\Gamma_{inh}$ is the total inhomogeneous width of the satellite line. This limits the optical depth of the prepared ensemble, which is an important concern for a photonic states stored with memory protocols, since a high optical depth is required to get a high efficiency recall. However, very high quantum memory efficiencies have been achieved using prepared spectral features containing parts-per-billion levels of rare earth ions. For example, an efficiency of 76\% was achieved in a 0.005\% Pr:Y$_2$SiO$_5$ crystal with an inhomogeneous linewidth of 3~GHz, corresponding to a concentration for a 1 MHz wide feature of 17 parts-per-billion (ppb) \cite{hedges10}. If we assume a higher threshold of 100 ppb to allow for higher efficiency and account for the lower oscillator strength of Eu compared to Pr, we can estimate the number of memory qubits possible for photonic processing applications. In a crystal with an intrinsic linewidth of 10 MHz and a dopant concentration of 0.1\% (contributing 10 MHz of extra broadening, expected in, for example,  \euform), three qubits could be produced with concentrations above the 100 ppb threshold.  Linewidths of 10 MHz have been observed in lightly doped Y$^7$LiF$_4$ \cite{macfarlane98, thiel11} and we have seen a linewidth of 25 MHz in the stoichiometric crystal \isoeuform \cite{ahlefeldt16}.  It is, therefore, feasible to construct systems with at least two qubits with good optical depth in existing materials. This allows operations on a single photonic qubit (encoded in two spin qubits). Achieving more spin qubits while maintaining a high optical depth will require reducing the linewidth down to the 1~MHz level where distillation is no longer necessary. If distillation is not used, the number of qubits is only limited by the number of resolvable satellite lines, which, in narrow materials, can be expected to be over the $\mathcal{O}$(30) lines seen in current materials \citeit{ahlefeldt13method, yamaguchi99}.

If the ensemble qubit system were used for quantum computing rather than photonic state processing, larger numbers of qubits are possible since the optical depth can be low and the ensemble small. Detection of ensembles containing $\mathcal{O}$($10^4$) ions with good signal-to-noise ratios is possible with fairly conventional optical setups \citeit{kroll93}, which means five to seven qubit systems could be created in existing materials. To see this, consider  a crystal with a 250 \AA$^3$ volume per rare earth atom (e.g. \euform), an 0.1\% doping concentration, a 5~mm length, and a 100~$\mu$m diameter region interacting with the laser drive fields. Then, an ensemble size of $10^4$ can be achieved for five qubits in crystals with linewidths below 100~MHz (e.g. \cite{ahlefeldt09}), while this size can be reached for seven qubits with a linewidth below 60~MHz (e.g. \cite{ahlefeldt16}).  Much smaller ensembles could be detected if the techniques developed for single rare earth ion detection \citeit{kolesov12, zhong18} were employed.

Both applications we consider require low optical inhomogeneous linewidths.  For computing, an additional priority is having low-error single and two-qubit gates, particularly if the system is used for error correcting, since we want the error induced by the error correction sequence to be smaller than the errors being corrected. For photonic processing, low-error gates are desired to minimize corruption of the photonic state, but the system must also serve as a good quantum memory. These two requirements, a good memory system and low-error gates, are dependent on several material properties, as explained below.

For a good memory, the system must have a long hyperfine lifetime (minutes to hours)  to allow efficient optical pumping preparation of the qubit. A long hyperfine coherence time is also required to obtain a long memory storage time, needed for most network applications of quantum memories and photonic processing. The memory storage time is limited by the hyperfine decoherence time, and interactions  with the fluctuating nuclear spin bath in the crystal are typically the dominant decoherence mechanism. The coherence time can be improved using magnetic fields. A small field in any direction will slow down spin flips and reduce the coupling to the rare earth ion. We can also use the zero first order Zeeman (ZEFOZ) technique, where a field is applied along a specific direction to almost entirely turn off the coupling to the spin bath, to further extend the coherence time.

Meanwhile, the error in a computing gate is determined by the errors in the individual pulses making up the gate. Errors in driving, for example, an optical $\pi$ pulse come from three main sources: off-resonant excitation of other transitions in the system, optical decoherence during the pulse, and  errors in the end state because the inhomogeneously broadened ensemble is not all driven to the same state (the pulse is not completely ``hard''). These error sources suggest a system with large hyperfine splittings to minimize off-resonant excitation and a long coherence time relative to the  optical Rabi frequency to minimize decoherence. The Rabi frequency itself is also determined by the sources of error, and there will be an optimal value for any system. At low Rabi frequencies, inhomogeneous driving errors occur, and gates are slow so errors due to decoherence are large. At high Rabi frequencies errors due to off-resonant excitation dominate.

To further reduce error, we want an ``ideal'' $\Lambda$ system: the oscillator strength from $\ket{e}$ is 50\% in the   $\ket{0}\rightarrow \ket{e}$ transition and 50\%  in  the $\ket{1}\rightarrow\ket{e}$ transition, with no oscillator strength to any other transitions in the system. This is not common in rare earth optical transitions at zero field, but oscillator strengths can be manipulated to be closer to an ideal $\Lambda$ transition by applying magnetic fields \citeit{goldner04,goldner06a}. 

So far this section has described generally what properties are desired of the host ion and host crystal. We will finish by briefly evaluating different materials.

For a host ion, among the non-Kramers rare earth ions, Pr\tplus and Eu\tplus offer good coherence times and sufficient hyperfine structure, but Eu\tplus has the larger hyperfine splittings, suggesting it is the better candidate. Further, because both states in the optical $^5$D$_0-^7$F$_0$ are $J=0$, it has a low sensitivity to the crystalline environment and thus a lower inhomogeneous linewidth than Pr\tplus in most materials. The weaker oscillator strength of this $0\leftrightarrow0$ optical transition does mean that an Eu\tplus crystal will have lower optical depth and Rabi frequency for fixed optical power than an equivalent Pr\tplus crystal.

Kramers ions offer much larger hyperfine structure than non-Kramers ions, but typically have hyperfine lifetimes and coherence times of milliseconds or less due to interactions with the electron spin.  However, we recently showed that a large magnetic field and low temperature can extend the hyperfine lifetime of Er\tplus to 10 minutes and coherence time to 1 second in \yso by freezing the electron spin \citeit{rancic18}. In this regime Kramers ions are very attractive for photonic applications, particularly Er\tplus since it has an optical transition in the 1550 nm telecommunications band and only a single isotope with hyperfine structure. However, the effect of placing a Kramers ion, with its large electronic  magnetic moment, stoichiometric in a crystal is not well studied. We do not yet know if it is possible to create a system with only (diagonal) frequency shift interactions between neighboring qubits.

When choosing the host crystal for these ions, the most restrictive of the above criteria is narrow optical inhomogeneous linewidths, since linewidths below 10 MHz are likely required to make systems of more than three qubits with high optical depth. Isotopically purified \isoeuform is the only stoichiometric material to have shown a linewidth near 10 MHz. This material also has coherence times similar to lightly doped Eu\tplus crystals ($\mathcal{O}$(10 ms) on hyperfine transitions and $\mathcal{O}$(1 ms) on optical transitions) when deuterated to remove the dual effects of magnetic noise due to the large moment of H and non-radiative decay of the optical excited state via high energy O-H phonons \cite{ahlefeldt13optical}. These properties make it the most promising material identified so far. There is nothing particularly special about \euform, though, and  lower linewidth materials should exist.  Initial measurements we have made in other hydrated crystals suggest similar linewidths are possible. It is also worth investigating growing high quality non-hydrated stoichiometric crystals, since hydrated materials are not suitable for Er\tplus or most other rare earth ions, where the smaller optical energy gaps mean they will be more strongly affected by non-radiative decay than Eu\tplus. 

Another way of reaching larger numbers of qubits is to be able to detect smaller ensembles. The ultimate aim would be to detect single-instance spin clusters, at which point the scaling argument becomes very different. The inhomogeneous broadening that limited ensemble qubits is absent in single instance. Instead, the optical frequency resolution of the qubits becomes the limit. 

We will use \euformd as an example to estimate the maximum number of qubits for a single-instance spin cluster. To consider two qubits resolved, we need that gates applied to one qubit do not cause errors on the other qubit. Practically, the easiest way to ensure this is to choose a separation similar to the hyperfine splitting of the computing levels' optical transitions. In \euformd, this is of the order of 50~MHz.

Estimating the number of satellite lines in the crystal separated by this amount is more difficult, because the interaction causing satellite structure in rare earth crystals has not been identified and its scaling with separation is unknown. Possible interactions include electric dipole-dipole, electric multipole, and superexchange. At nearest neighbour distances where we have measured interactions, the latter two are important, but the dipole interaction will dominate at larger distances, and thus is the most important interaction for the majority of satellite lines in a single instance system. We can roughly estimate the number of resolvable lines by  assuming that  10\% (1GHz) of the $\mathcal{O}$(10 GHz) shift seen for nearest-neighbour satellite lines in \euformd is due to an electric dipole-dipole interaction. Since the dipole-dipole interaction decays as $\frac{1}{r^3}$, sites out to 20\AA\ will have satellite line shifts greater than 50 MHz, corresponding to about 100 sites. The C$_2$ symmetry of \euformd means these correspond to only around 50 satellite lines, but all sites will be accessible in an equivalent C$_1$ crystal.

In a computer that size, every qubit will not have a sufficient interaction with every other qubit to allow direct two qubit gates. However, the qubits are still strongly interconnected. For the interaction between two qubits to be resolved, we require it to be larger than the optical Rabi frequency. In single instance, a Rabi frequency of 1~MHz gives low error gates, so we estimate that interactions of around 10~MHz are required. To work out how many ions have this size interaction, we need to know how the optical frequency shift interaction scales with distance. Interactions between pairs of ions in \euform ranging from 0.5-46 MHz have been measured, but the separation of the pair could only be determined for some cases \citeit{ahlefeldt13precision}. For example, the 46~MHz interaction was between ions with a separation no smaller than 7.9~\AA, while a 1.7 MHz interaction was likely to be due to ions separated by 30~\AA. From these numbers, it is reasonable to estimate that 10~MHz interactions are possible between ions separated by around 10~\AA, equating to 10 qubits. Thus, every qubit in a single instance system would be well-connected to around 10 others.

A fully scalable quantum computer can be created using these single instance spin clusters if strong coupling between multiple spin clusters via an optical field can be achieved.  A path towards achieving this coupling is the approach of Zhong et al. \citeit{zhong18}: a photonic crystal resonator created in the surface of a rare earth crystal is used to generate a high coupling strength between light and single rare earth ions present in the resonator.

\section{Conclusion}
We described a scheme for making small qubit systems in stoichiometric rare earth crystals using optical-frequency-addressed spin clusters created by doping the crystal with a substitutional defect. In contrast to previous quantum computing proposals using spin clusters, such as using NV-\iso{13}C clusters or doped rare earth crystals, the cluster is formed from host ions at nearest-neighbor separations, so each cluster in the crystal is identical and the qubit-qubit interactions used for computing gates are strong and homogeneous. The scheme is applicable to both single-instance and ensemble implementations, although at the moment detection of single rare earth ions is challenging. 
 We discussed how small ensemble quantum computers in this system might be used to study the effect of error correlation on error correction protocols. We also showed that ensemble qubit systems can be used to perform linear processing of photonic states stored as frequency-encoded dual-rail qubits across two ensemble spin qubits in the system. Both single qubit and two-qubit operations on the photonic states can be achieved using SWAP gates applied on pairs of ensemble qubits. This approach to linear optics quantum computing has the advantage that it avoids multiple recalls of the state from the quantum memory, thus leading to a higher efficiency computation. 
 
\begin{acknowledgements}
The authors wish to thank N. Menicucci for helpful discussions. RLA is a recipient of an Australian Research Council Discovery Early Career Researcher Award (project No. DE170100099). This work was supported by the Australian Research Council Centre of Excellence for Quantum Computation and Communication Technology (Grant No. CE110001027).  MRA is now an employee of the quantum control company Q-CTRL. His contribution to this work was completed entirely during his time at UNSW Canberra.
\end{acknowledgements}

\bibliography{stochiometric_bib}

 \end{document}